\documentclass[12pt]{cernart}
\usepackage{t1enc}
\usepackage{times,euler}
\usepackage{amsmath,amsfonts,amssymb}
\usepackage[latin2]{inputenc}
\usepackage{color}
\usepackage{graphicx}

\newcommand{\be}{\begin{eqnarray}}
\newcommand{\ee}{\end{eqnarray}}
\newcommand{\nn}{\nonumber}

\newcommand{\bi}{\begin{itemize}}
\newcommand{\ei}{\end{itemize}}
\newcommand{\bc}{\begin{center}}
\newcommand{\ec}{\end{center}}
\newcommand{\la}{\langle}
\newcommand{\ra}{\rangle}

\begin{document}
\begin{titlepage}
\docnum{CERN-PH-EP/2004-013}
\date{5th April 2004}
\title{
 \bf\Large Measurement of the Branching Ratio and Form Factors for the Decay \boldmath $K_L\rightarrow\pi^{\pm}\pi^0e^{\mp}\nu_e(\bar\nu_e)$
}
\begin{Authlist}

\vspace{1cm}

{\large NA48 Collaboration} \\

\vspace{1cm}

\begin{center}
J.R.~Batley,
R.S.~Dosanjh,
T.J.~Gershon,
G.E.~Kalmus,
C.~Lazzeroni,
D.J.~Munday,
E.~Olaiya,
M.~Patel,
M.A.~Parker,
T.O.~White,
S.A.~Wotton \\
{\em Cavendish Laboratory, University of Cambridge, Cambridge, CB3 0HE, U.K.\footnotemark[1]} \\[0.2cm] 
R.~Arcidiacono,
G.~Bocquet,
A.~Ceccucci,
T.~Cuhadar-D\"{o}nszelmann,
D.~Cundy\footnotemark[2],
N.~Doble\footnotemark[3],
V.~Falaleev,
L.~Gatignon,
A.~Gonidec,
B.~Gorini,
P.~Grafstr\"om,
W.~Kubischta,
I.~Mikulec\footnotemark[4],
A.~Norton,
S.~Palestini,
H.~Wahl\footnotemark[5] \\
{\em CERN, CH-1211 Gen\`eve 23, Switzerland.} \\[0.2cm] 
C.~Cheshkov\footnotemark[6],
P.~Hristov\footnotemark[6],
V.~Kekelidze,
L.~Litov\footnotemark[6],
D.~Madigozhin,
N.~Molokanova,
Yu.~Potrebenikov,
A.~Zinchenko \\
{\em Joint Institute for Nuclear Research, Dubna, 141980, Russian Federation} \\[0.2cm] 
P.~Rubin\footnotemark[7],
R.~Sacco\footnotemark[8],
A.~Walker \\
{\em Department of Physics and Astronomy, University of Edinburgh, Edinburgh, EH9 3JZ, U.K.\footnotemark[1]} \\[0.2cm] 
D.~Bettoni,
R.~Calabrese,
P.~Dalpiaz,
J.~Duclos,
P.L.~Frabetti\footnotemark[9],
A.~Gianoli,
M.~Martini,
L.~Masetti\footnotemark[10],
F.~Petrucci,
M.~Savri\'e,
M.~Scarpa \\
{\em Dipartimento di Fisica dell'Universit\`a e Sezione dell'INFN di Ferrara, I-44100 Ferrara, Italy} \\[0.2cm] 
A.~Bizzeti\footnotemark[11],
M.~Calvetti,
G.~Collazuol\footnotemark[3],
E.~Iacopini,
M.~Lenti,
F.~Martelli\footnotemark[12],
G.~Ruggiero,
M.~Veltri\footnotemark[12] \\
{\em Dipartimento di Fisica dell'Universit\`a e Sezione dell'INFN di Firenze, I-50125 Firenze, Italy} \\[0.2cm] 
%
%
K.~Eppard,
M.~Eppard\footnotemark[6],
A.~Hirstius\footnotemark[6],
K.~Kleinknecht,
U.~Koch,
L.~K\"opke,
P.~Lopes~da~Silva,
P.~Marouelli,
I.~Mestvirishvili,
C.~Morales,
I.~Pellmann\footnotemark[13],
A.~Peters\footnotemark[6],
B.~Renk,
S.A.~Schmidt,
V.~Sch\"{o}nharting,
R.~Wanke,
A.~Winhart\\
{\em Institut f\"ur Physik, Universit\"at Mainz, D-55099 Mainz, Germany\footnotemark[15]} \\[0.2cm] 
J.C.~Chollet,
L.~Fayard,
G.~Graziani,
L.~Iconomidou-Fayard,
G.~Unal,
I.~Wingerter-Seez \\
{\em Laboratoire de l'Acc\'el\'erateur Lin\'eaire, IN2P3-CNRS, Universit\'e de Paris-Sud, F-91898 Orsay, France\footnotemark[16]} \\[0.2cm] 
G.~Anzivino,
P.~Cenci,
E.~Imbergamo,
G.~Lamanna,
P.~Lubrano,
A.~Mestvirishvili,
A.~Michetti,
A.~Nappi,
M.~Pepe,
M.~Piccini,
M.~Valdata-Nappi \\
{\em Dipartimento di Fisica dell'Universit\`a e Sezione dell'INFN di Perugia, I-06100 Perugia, Italy} \\[0.2cm] 
R.~Casali,
C.~Cerri,
M.~Cirilli\footnotemark[6],
F.~Costantini,
R.~Fantechi,
L.~Fiorini,
S.~Giudici,
I.~Mannelli, 
G.~Pierazzini,
M.~Sozzi \\
{\em Dipartimento di Fisica, Scuola Normale Superiore e Sezione dell'INFN di Pisa, I-56100 Pisa, Italy} \\[0.2cm]   
J.B.~Cheze,
M.~De Beer,
P.~Debu,
F.~Derue,
A.~Formica,
G.~Gouge,
G.~Marel,
E.~Mazzucato,
B.~Peyaud,
R.~Turlay\footnotemark[14],
B.~Vallage \\
{\em DSM/DAPNIA - CEA Saclay, F-91191 Gif-sur-Yvette, France} \\[0.2cm] 
M.~Holder,
A.~Maier,
M.~Zi\'o\l kowski \\
{\em Fachbereich Physik, Universit\"at Siegen, D-57068 Siegen, Germany\footnotemark[17]} \\[0.2cm] 
C.~Biino,
N.~Cartiglia,
M.~Clemencic,
F.~Marchetto, 
E.~Menichetti,
N.~Pastrone \\
{\em Dipartimento di Fisica Sperimentale dell'Universit\`a e Sezione dell'INFN di Torino, I-10125 Torino, Italy} \\[0.2cm] 
J.~Nassalski,
E.~Rondio,
W.~Wi\'slicki,
S.~Wronka \\
{\em Soltan Institute for Nuclear Studies, Laboratory for High Energy Physics, PL-00-681 Warsaw, Poland\footnotemark[18]} \\[0.2cm] 
H.~Dibon,
M.~Jeitler,
M.~Markytan,
G.~Neuhofer,
M.~Pernicka,
A.~Taurok,
L.~Widhalm \\
{\em \"Osterreichische Akademie der Wissenschaften, Institut f\"ur Hochenergiephysik, A-1050 Wien, Austria\footnotemark[19]} 
\end{center}
\end{Authlist}

\begin{center}

\vspace{1cm}

Submitted{ to Physics Letters B}

\vspace{1cm}

\abstract{
The $K_L\rightarrow\pi^{\pm}\pi^0e^{\mp}\nu_e(\bar\nu_e)$ decay was investigated with the NA48 detector at CERN SPS using a beam of long-lived neutral kaons. 
The branching ratio $Br(K_L\rightarrow\pi^{\pm}\pi^0e^{\mp}\nu_e(\bar\nu_e))=(5.21\pm 0.07_{stat}\pm 0.09_{syst})\times 10^{-5}$ was fixed from a sample of 5464 events with 62 background events.
The form factors $\bar{f}_s$, $\bar{f}_p$, $\lambda_g$ and $\bar{h}$ were found to be in agreement with previous measurements but with higher accuracy.
The coupling parameter of the chiral Lagrangian $L_3=(-4.1\pm 0.2)\times 10^{-3}$ was evaluated from the data.
}

\end{center}

\maketitle

\footnotetext[1]{Funded by the U.K.\ Particle Physics and Astronomy Research Council.}
\footnotetext[2]{Present address: Istituto di Cosmogeofisica del CNR di Torino, I-10133 Torino, Italy.}
\footnotetext[3]{Also at Dipartimento di Fisica dell'Universit\`a e Sezione dell'INFN di Pisa, I-56100 Pisa, Italy.}
\footnotetext[4]{On leave from \"Osterreichische Akademie der Wissenschaften, Institut f\"ur Hochenergiephysik, A-1050 Wien, Austria.}
\footnotetext[5]{Present address: Dipartimento di Fisica dell'Universit\`a e Sezione dell'INFN di Ferrara, I-44100 Ferrara, Italy.}
\footnotetext[6]{Present address: CERN, CH-1211 Gen\`eve 23, Switzerland.} 
\footnotetext[7]{Permanent address: Department of Physics and Astronomy, George Mason University, Fairfax, VA 22030, USA, supported in part by the US NSF under award \#0140230.}
\footnotetext[8]{Present address: Laboratoire de l'Acc\'el\'erateur Lin\'eaire, IN2P3-CNRS,Universit\'e de Paris-Sud,
                       F-91898 Orsay, France.}
\footnotetext[9]{Present address: Joint Institute for Nuclear Research, Dubna, 141980, Russian Federation.}
\footnotetext[10]{Present address: Institut f\"ur Physik, Universit\"at Mainz, D-55099 Mainz, Germany.}
\footnotetext[11]{Dipartimento di Fisica dell'Universita' di Modena e Reggio Emilia, via G.~Campi 213/A, I-41100, Modena, Italy.}
\footnotetext[12]{Instituto di Fisica Universit\'a di Urbino, I-61029 Urbino, Italy.}
\footnotetext[13]{Present address: DESY Hamburg, D-22607 Hamburg, Germany.}
\footnotetext[14]{Deceased.}
\footnotetext[15]{Funded by the German Federal Minister for Research and Technology (BMBF) under contract 7MZ18P(4)-TP2.}
\footnotetext[16]{Funded by Institut National de Physique des Particules et de Physique Nucl\'eaire (IN2P3), France}
\footnotetext[17]{Funded by the German Federal Minister for Research and Technology (BMBF) under contract 056SI74.}
\footnotetext[18]{Supported by the Committee for Scientific Research grants 5P03B10120 and SPUB-M/CERN/P03/DZ210/2000.}
\footnotetext[19]{Funded by the Austrian Ministry of Education, Science and Culture under contract GZ 616.360/2-IV GZ
                  616.363/2-VIII, and by the Fund for Promotion of Scientific Research in Austria (FWF) under contract P08929-PHY.}

\end{titlepage}

\section{Introduction}

The decay $K_L\rightarrow\pi\pi e\nu$, called $K_{e4}$, is recognized as a good test for chiral perturbation theory (CHPT) and its predictions for long-distance meson interactions.
In particular, it is used to determine the $\pi\pi$ partial wave expansion parameters: threshold parameters, slopes and scattering lengths, where the S-wave $\pi\pi$ scattering lengths can be further related to the quark condensate \cite{colangelo}.
The complete set of CHPT parameters has been calculated in the one-loop approximation ${\cal O}(p^4)$ and the form factors $F$ and $G$ and quark condensates in the two-loop approximation ${\cal O}(p^6)$ \cite{bijnens}.

Following the initial observation of charged $K_{e4}$ \cite{chargedlow}, the process $K^+\rightarrow\pi^+\pi^- e^+\bar\nu_e$, called $K^+_{e4}$, was measured in ref. \cite{rosselet} based on an event sample of 30,000 events and, more recently, a high-statistics experiment \cite{pislak} detected $400,000$ such decays. 
Those experiments determined the $K^+_{e4}$ decay rate, four form factors and the difference of the $s$- and $p$-wave phase shifts $\delta_0^0-\delta_1^1$ as a function of the mass of the pion pair $M_{\pi\pi}$.
By fitting the Roy model \cite{roy} to the $M_{\pi\pi}$-dependence of phase shifts and assuming time-reversal invariance, they also evaluated the scattering length $a_0^0$.

After a low-statistics observation of the neutral $K_{e4}$ decay $K_L\rightarrow\pi^{\pm}\pi^0 e^{\mp}\nu_e(\bar{\nu}_e)$ \cite{carrol}, a more complete analysis was performed in ref. \cite{makoff}, where a sample of 729 events was used to determine the branching ratio, the threshold value $g(M_{\pi\pi}=0)$ of the $g$ form factor, the relative form factors $\bar{f}_s=f_s/g$, $\bar{f}_p=f_p/g$ and $\bar{h}=h/g$, and the $M_{\pi\pi}$-dependence of $g$.
The neutral $K_{e4}$ decay is well-suited for measuring the $G$ form factor and the $L_3$ parameter of CHPT. 

This paper reports on the measurement of both the branching ratio and the form factors of neutral $K_{e4}$ decays by the NA48 Collaboration at CERN, using a significantly larger data sample than previous measurements.
In addition, the coefficient $L_3$ of the chiral Lagrangian, sensitive to the gluon condensate, is evaluated with high accuracy.

\section{Kinematics and parametrization of the decay cross section}

The matrix element for the decay is assumed to factorize into a leptonic term, describing the coupling of the $W$ boson to leptons, and an hadronic term, accounting for hadronization of quarks into pions and representing the $V-A$ structure \cite{treiman}
\be
M=\frac{G_F}{\sqrt{2}}\sin\Theta_C\langle\pi\pi|A^{\lambda}+V^{\lambda}|K\rangle \bar u_{\nu}\gamma_{\lambda}(1-\gamma_5)\nu_e,
\label{eq01}
\ee
where $G_F$ is the Fermi weak coupling constant and $\Theta_C$ the Cabibbo angle.
The vector component $\la \pi\pi|V|K\ra$ is parametrized in terms of one form factor $H$
\be
\langle\pi\pi|V^{\lambda}|K\rangle =\frac{1}{m_K^3}H\varepsilon^{\lambda\mu\nu\rho}(p_K)_\mu(p_{\pi_1}+p_{\pi_2})_{\nu}(p_{\pi_1}-p_{\pi_2})_{\rho},
\label{eq02}
\ee
and the axial-vector part $\la \pi\pi|A|K\ra$ in terms of three form factors: $F$, $G$ and $R$
\be
\langle\pi\pi|A^{\lambda}|K\rangle =\frac{1}{m_K}[F(p_{\pi_1}+p_{\pi_2})+G(p_{\pi_1}-p_{\pi_2})+R(p_K-p_{\pi_1}-p_{\pi_2})]^{\lambda},
\label{eq03}
\ee
where the $R$ term is suppressed by the squared ratio of the electron mass to the kaon mass and can therefore be neglected.

\begin{figure}[h]
\begin{center}
\includegraphics[width=100mm,height=50mm]{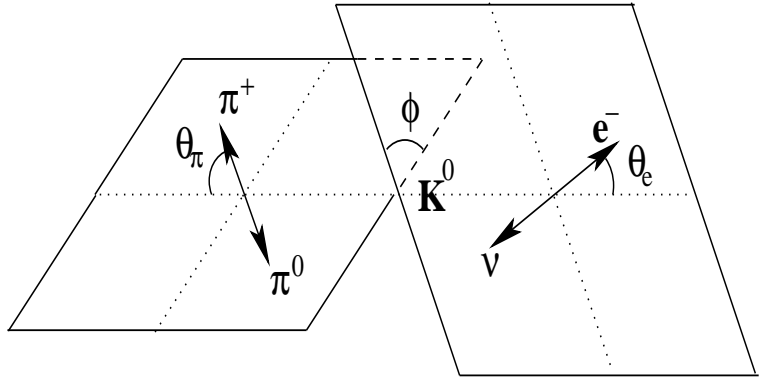}
\caption{Definition of the Cabibbo-Maksymowicz kinematic variables \cite{cabibbo} used for the analysis of $K_{e4}$ decays. The angles $\theta_{\pi}$ ($\theta_e$) are between the charged pion (electron) momentum in the dipion (dilepton) centre-of-momentum frame and the dipion (dilepton) momentum in the kaon rest frame. The directed angle $\phi$ is from the $\pi\pi$ plane to the $e\nu$ plane.}
\label{fig1}
\end{center}
\end{figure}

 The differential cross section for the $K_{e4}$ decay was proposed \cite{cabibbo,treiman} to be analysed in terms of five Cabibbo-Maksymowicz (C-M) variables: invariant masses of the dipion $M_{\pi\pi}$ and the dilepton $M_{e\nu}$, the polar angles $\theta_{\pi}$ ($\theta_e$) between the charged pion (electron) momentum in the $\pi\pi$ ($e\nu$) centre-of-momentum frame and the dipion (dilepton) momentum in the kaon rest frame, and the azimuthal angle $\phi$ from the $\pi\pi$ plane to the $e\nu$ plane (Fig. \ref{fig1}).

The $\theta_{\pi}$-dependence of the form-factors is made explicit by using a partial-wave expansion of the hadronic matrix element with respect to the angular momentum of the pion pair and restricting this expansion to $s$ and $p$ waves due to the limited phase space available in the $K_{e4}$:
\be
F & = & f_s e^{i\delta_s}+f_p e^{i\delta_p}\cos \theta_{\pi} \nn \\
G & = & g e^{i\delta_p} \nn \\
H & = & h e^{i\delta_p}
\label{eq1}
\ee
The $G$ and $H$ expansions contain only the $p$ wave due to their antisymmetry with respect to pion exchange.
Using the partial wave decomposition (\ref{eq1}), an explicit expansion of the decay cross section in terms of the form factors $f_s$, $f_p$, $g$, $h$ and $\delta=\delta_s-\delta_p$, and C-M variables $M_{\pi\pi}$, $M_{e\nu}$, $\cos\theta_{\pi}$, $\cos\theta_e$ and $\phi$, was taken from ref. \cite{rosselet}.
The possible $M_{\pi\pi}$-dependence of $g$ was accounted for by parameterizing $g(M_{\pi\pi})=g(0)[1+\lambda_g(M_{\pi\pi}^2/4m_{\pi}^2-1)]$, where $\lambda_g$, together with the form factors, has to be determined from a fit to the data, and $m_{\pi}$ stands for the average of the charged and neutral pion masses.

\section{The beam}

The NA48 experiment used for this investigation a 400 GeV/c proton beam from the CERN Super Proton Synchroton with a nominal intensity of $1.5\times 10^{12}$ protons per spill, delivered every 16.8 s in 4.8 s long spills \cite{biino}.
Two kaon beams, one providing $K_L$ decays and called the $K_L$ beam, and another one, providing $K_S$ decays, and called the $K_S$ beam, were produced simultaneously on two separate targets.
For the $K_{e4}$ measurement only the $K_L$ beam was relevant.
The $K_L$ beryllium target was located 126~m before the decay region.
Charged particles were swept by dipole magnets, and the remaining neutral beam was defined by a set of collimators.
The total flux of $K_L$'s at the entrance of the fiducial decay volume was $2\times 10^7$ per spill.

\section{The detector}

The detector system, located 114~m after the $K_S$ target and extending 35~m downstream, consisted of two principal subsystems: a magnetic spectrometer and a spectrometer for neutral decays.
In addition, there were scintillating hodoscopes, a hadron calorimeter, muon veto counters, beam veto counters, and a tagging station on the $K_S$ beamline.

The magnetic spectrometer was contained in a helium tank and consisted of a dipole magnet with a transverse momentum kick of 265 MeV/$c$ and four drift chambers,
each equipped with eight sensitive planes, arranged two before and two after the magnet.
The momentum resolution of this spectrometer was between $0.5\,\%$ and $1\,\%$, depending on the momentum, and the average plane efficiency exceeded $99\,\%$.

A scintillating hodoscope, consisting of two orthogonal planes of scintillating strips (horizontal and vertical) had a time resolution of 150 ps. 
Signals from quadrants were logically combined and used for triggering charged events in the first level trigger.

An iron-scintillator hadron calorimeter, 6.7 nuclear interactions thick and located downstream of both spectrometers, provided a total energy measurement, complementary to the electromagnetic calorimeter.

Muon veto counters, situated behind the hadron calorimeter, provided time information used to identify muons and to suppress backgrounds both in low-level triggers and offline.

The fiducial decay region was surrounded by seven sets of iron-plastic veto scintillators, called AKL, used for identification of photons escaping this volume.

The spectrometer for neutral decays consisted of a quasi-homogeneous ionization chamber calorimeter filled with 10 m$^3$ of liquid Krypton.
Its length, amounting to 27 radiation lengths with a Moli\`ere radius of 4.7 cm, ensured full containment of electromagnetic shower of energies up to 100 GeV,
excluding detector regions close to the edges.
The calorimeter was divided into 13,212 cells, $2~{\rm cm}\times 2 $ cm transversally to the beam, read out individually.
This calorimeter provided the resolution of reconstructed energy $\sigma(E)/E=9\,\%/E\oplus 3.2\,\%/\sqrt{E}\oplus 0.42\,\%$ and good reconstruction of the neutral vertex position along the beam.
Signals from the calorimeter were digitized asynchronously by a 40 MHz flash ADCs and read out with online zero-suppression.

A more detailed description of the apparatus can be found in ref. \cite{na48CP}.

\section{The trigger}

Data were taken using the minimum-bias trigger ETOT, requiring a minimal energy deposit of 35 GeV in the calorimeters, hit multiplicity in the first drift chamber, and a coincidence between opposite quadrants of the scintillator hodoscope. 
Since this trigger was downscaled by a factor of 30, a dedicated trigger KE4
was added to enhance statistics. 
The KE4 trigger used the neutral trigger system of NA48, which gave information about $x$ and $y$ projections of the energy deposited in the electromagnetic calorimeter. 
The requirement was at least 3, and not more than 5, clusters in any of the two projections, corresponding to the two photons from a $\pi^0$ decay and two charged particles, allowing for the loss of one cluster due to overlapping.
This trigger was downscaled by a factor of 50.
In the liquid crypton calorimeter readout the KE4 trigger induced lower threshold than the ETOT trigger.
In case of both triggers conditions were fulfilled, the lowest one was used.
Therefore a small number of events giving both triggers were included to the KE4 sample.  

The minimum-bias trigger ETOT was assumed to be fully efficient.
The efficiency of the KE4 trigger was measured relative to the ETOT trigger and found to be $(98.76\pm 0.21)\,\%$ for $K_{e4}$ events and  $(98.92\pm 0.01)\,\%$ for $K_{\pi 3}$ events.  The latter were used for normalization.

\section{Data sample}

The analysis described in this paper refers to the data collected in 2001.
The sample of $K_{e4}$ events was selected, from both the KE4 and ETOT triggers, by applying the following cuts to reconstructed events which fulfilled the triggering conditions:
\begin{enumerate}
\item Two well-reconstructed tracks of opposite charges.
\item Four reconstructed clusters in the liquid Krypton electromagnetic calorimeter.
\item Two photon clusters, each of energy between 3 and 100 GeV, not associated with the charged tracks.
\item Tracks impacting the Krypton calorimeter between 15~cm and 120~cm from the beam axis. 
This cut eliminates clusters close to the edge and thus not fully contained in the calorimeter.
\item A minimum distance between photon and the charged pion clusters of 15~cm, thus ensuring cleanliness of cluster reconstruction.
\item A minimum distance of 5~cm between a photon cluster and the extrapolation of the charged tracks from before the magnet, partially removing background from decays $K_L\rightarrow\pi^{\pm}e^{\mp}\nu_e$ with two additional photons, at least one of them coming either from internal or external bremsstrahlung ($K_{e3+2\gamma}$).
\item The total energy deposited in the Krypton calorimeter had to be larger than 30~GeV.
\item The energy over momentum ratio ($E/p$) for the electron candidate had to be larger than 0.9 and smaller than 1.1 and for the charged pion candidate smaller than 0.8 (cf. fig. \ref{fig2}).
\begin{figure}[h]
\begin{center}
\includegraphics[width=80mm,height=70mm]{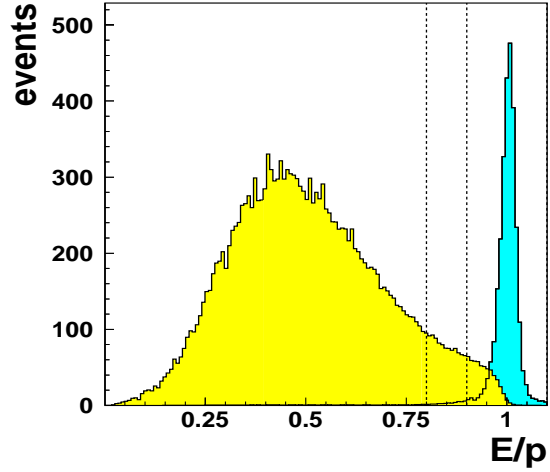} 
\caption{Distributions of $E/p$ for data. The wide distribution on the left corresponds to pions from cleanly selected $K_{\pi 3}$ sample, and the narrow one on the right to electrons from the $K_{e3}$ sample. These data were used to train the neural network. The cuts on $E/p$ are also shown.}
\label{fig2}
\end{center}
\end{figure}
\item A $\chi_{3\pi}^2$ variable for the $K_L\rightarrow\pi^+\pi^-\pi^0$ hypothesis was defined as
\be
\chi^2_{3\pi}=\Big(\frac{M_{3\pi}-M_K}{\sigma_M}\Big)^2+\Big(\frac{p_T-p_{T_0}}{\sigma_p}\Big)^2
\label{eq2}
\ee
with the invariant mass $M_{3\pi}$ under the $3\pi$ hypothesis and the transverse momentum $p_T$ (cf. fig. \ref{fig3}).  $M_K$ is the kaon mass, $p_{T_0}=0.006$ GeV/$c$ is the modal value of the $p_T$ distribution, $\sigma_M=0.0025$ GeV/$c$$^2$ and $\sigma_p=0.007$ GeV/$c$.
\begin{figure}
\begin{center}
\includegraphics[width=80mm,height=80mm]{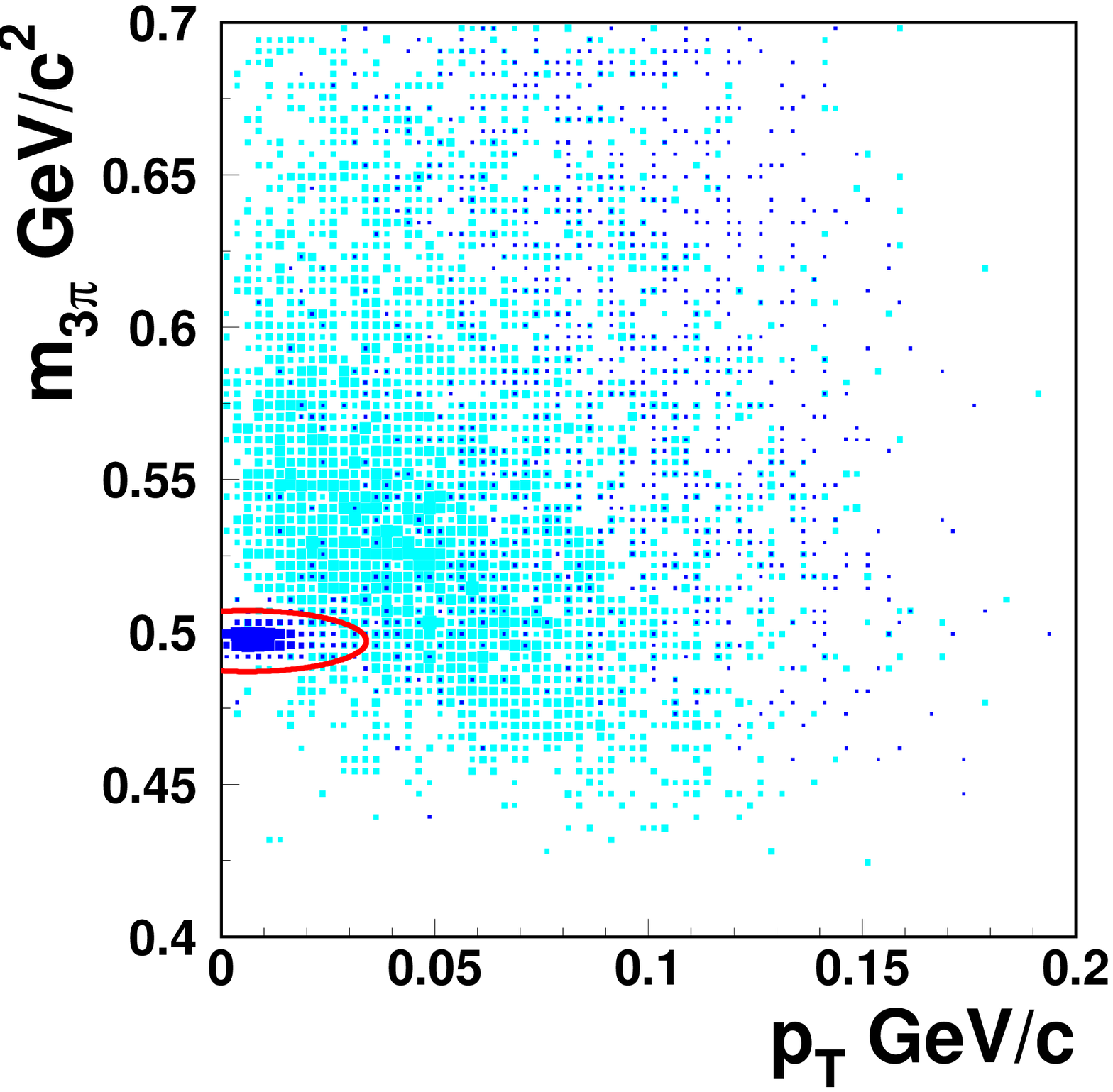} 
\caption{Distribution of measured mass of three visible particles, assuming the $3\pi$ hypothesis, versus their total $p_T$, for Monte Carlo $K_{e4}$ and $K_{\pi 3}$ events. The ellipse defines the cut $\chi^2_{3\pi}>16$ which distinguishes $K_{e4}$ events (outside the ellipse) from the $K_{\pi 3}$ background located around the kaon mass (inside the ellipse).}
\label{fig3}
\end{center}
\end{figure}
The cut $\chi^2_{3\pi}>16$ suppresses most of the $K_{\pi 3}$ background, where one of the charged pions is misidentified as the electron.
\item The invariant mass of the two-photon system, at the vertex defined by the two charged tracks, had to be between 0.11 and 0.15 GeV/$c$, which ensures that the photons come from a $\pi^0$ decay.
\item The ratio $p_T/E_{\nu}$ had to be between 0 and 0.02, where $p_T$ is the total transverse momentum of all visible particles (two pions and an electron) and $E_{\nu}$ is the energy of the neutrino in the laboratory frame.
This cut suppresses the $K_{e3+2\gamma}$ background with one or both photons coming from accidental coincidence. 
In this case the energy taken by the photons may be large enough to lead to a negative $E_{\nu}$ (cf. fig.~\ref{fig7}). 
The $K_{e3+2\gamma}$ event sample in fig.~\ref{fig7} was selected using a neural network algorithm.
The cut is not efficient for $K_{e3+2\gamma}$ background events with only bremsstrahlung photons, which are rejected by cuts 6 and 10.
\begin{figure}
\begin{center}
\includegraphics[width=80mm,height=80mm]{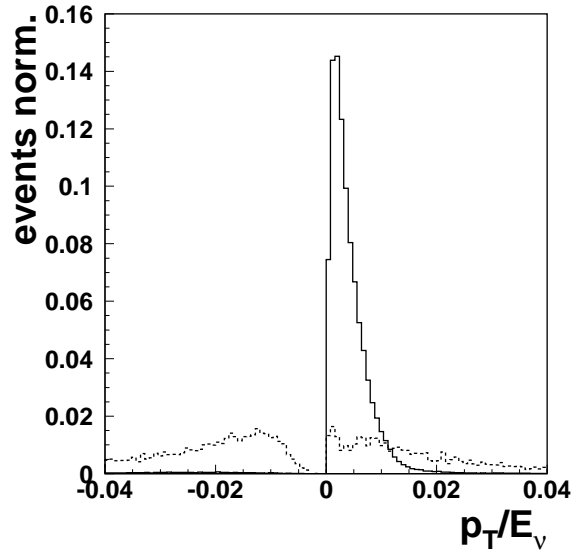}
\caption{Distributions of $p_T/E_{\nu}$ for the accepted and reconstructed Monte Carlo $K_{e4}$ events (solid) and the $K_{e3+2\gamma}$ background events from data (dashed). 
For $K_{e4}$, resolution smearing may occasionally lead to events with small negative values for $E_{\nu}$ and large negative $p_T/E_{\nu}$, which are hardly seen in the plot. 
The sharp edge of both distributions at $p_T/E_{\nu}=0$ is due to the kinematic suppression of large values of $E_{\nu}$.}
\label{fig7}
\end{center}
\end{figure}
\end{enumerate}

Cuts were also made on the maximum time difference between calorimeter clusters belonging to the same event and between clusters and tracks.  Cluster
quality criteria were met as were requirements for spatial cluster separation, cluster versus track spatial matching, and vertex position and quality.
Cuts 6, 10 and 11 above suppress background from $K_{e3+2\gamma}$ down to the level of 1.5\% and 2.2\% for the KE4 and ETOT triggers, respectively.
This was estimated from Monte Carlo by normalizing the $p_T/E_{\nu}$ spectrum from the $K_{e3+2\gamma}$ to the one from $K_{e4}$ in the region of $p_T/E_{\nu}<0$ and calculating the contamination for $p_T/E_{\nu}\ge 0$.

Requirements 8 and 9 eliminate most of the background from the $K_{\pi 3}$ channel.
\begin{figure}
\begin{center}
\begin{tabular}{cc}
\includegraphics[width=70mm,height=70mm]{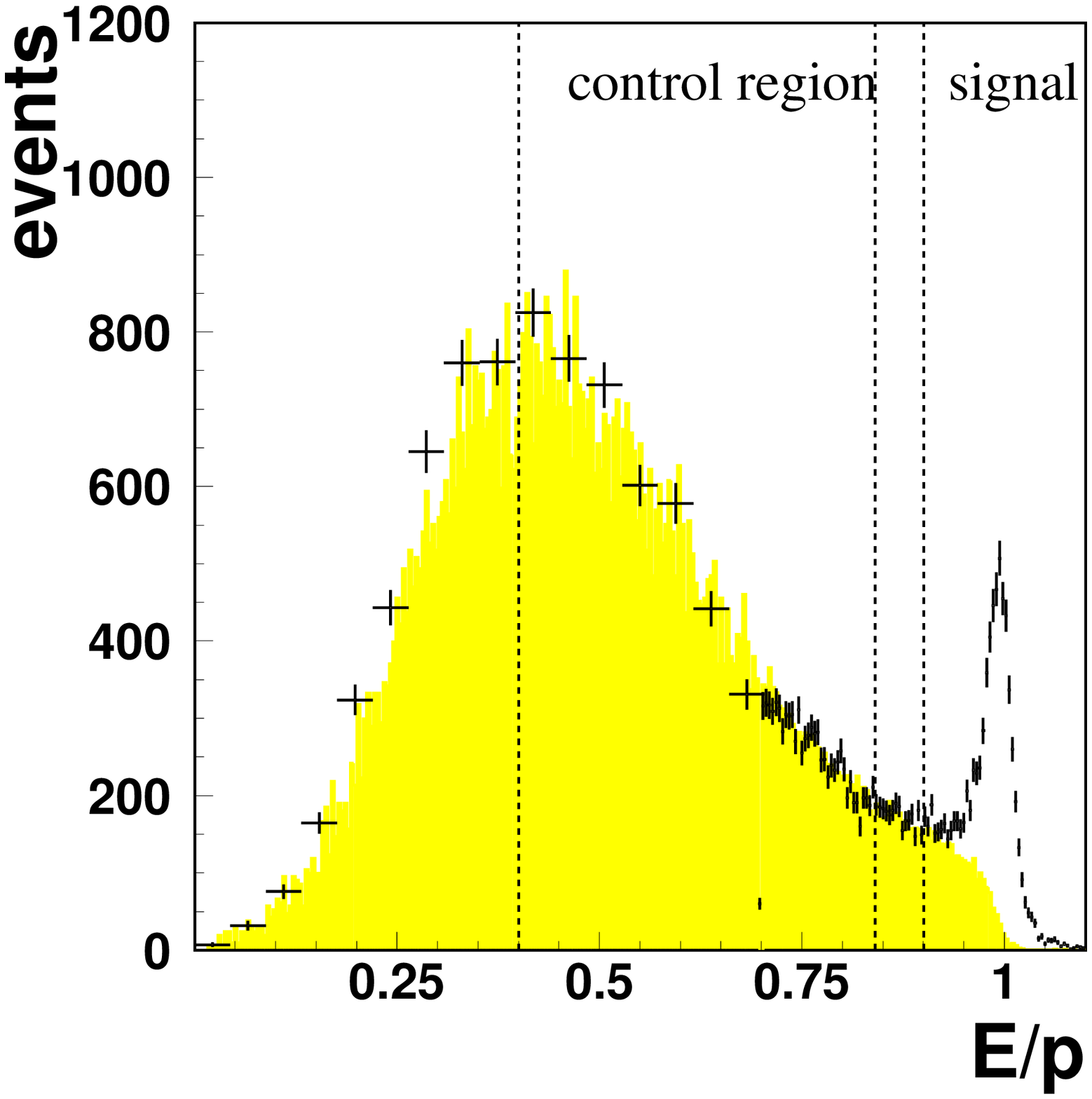} & \includegraphics[width=70mm,height=70mm]{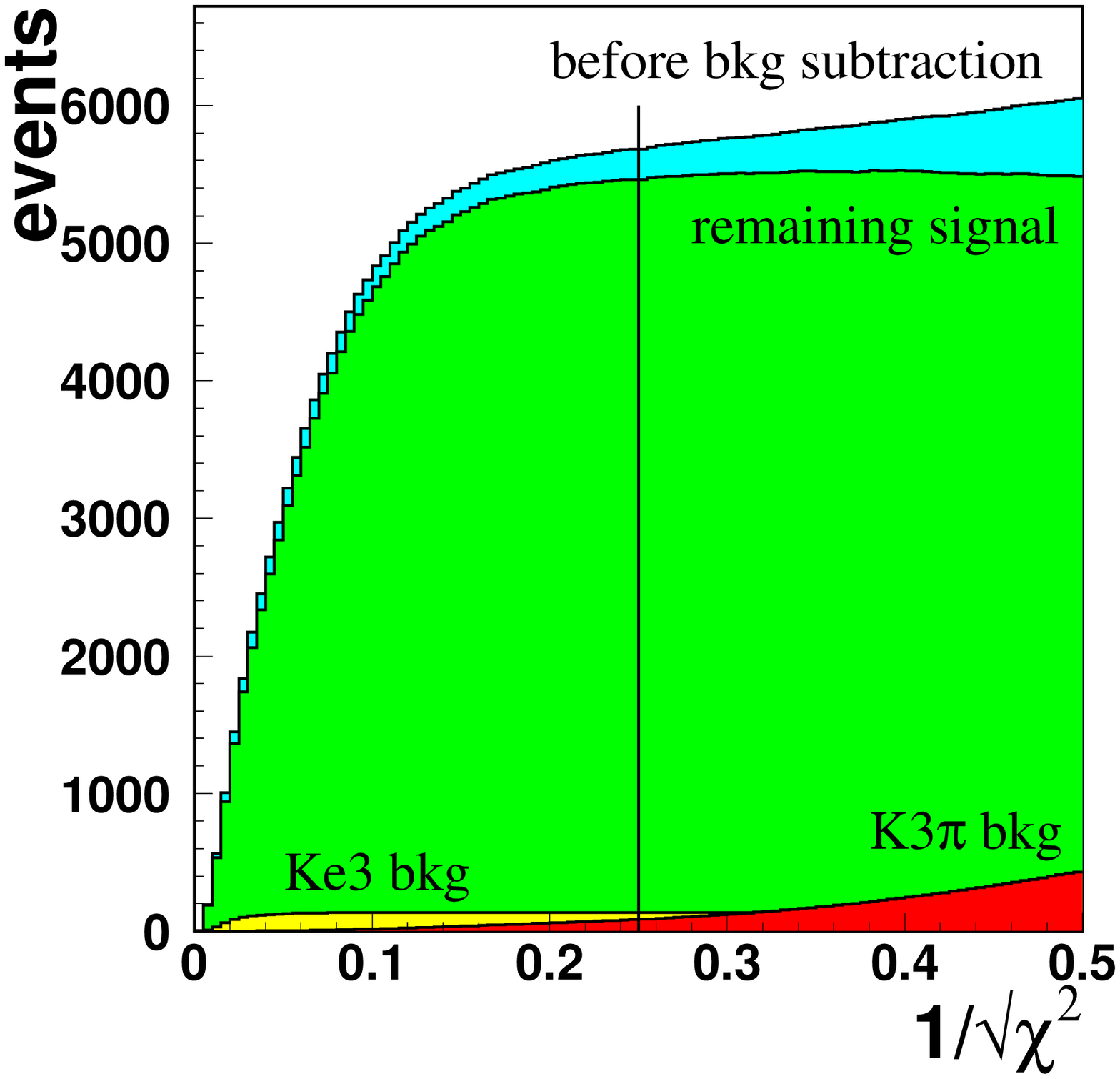} 
\end{tabular}
\caption{The $K_{\pi 3}$ background shown as functions of the $E/p$ with $\chi^2_{3\pi}>16$ (left) and of the $1/\chi^2_{3\pi}$ cut (right). For the $E/p$ distributions (left), the data were corrected for the downscaling factor when $E/p<0.7$ which results in larger errors in that domain. In the right-hand figure, the lower curves show the remaining $K_{e3}$ and $K_{\pi 3}$ backgrounds, the middle curve corresponds to the remaining signal, and the upper one shows all events. Both plots show data.}
\label{fig3p}
\end{center}
\end{figure}
The effects of the $\chi^2_{3\pi}$ and $E/p$ cuts are illustrated in figs \ref{fig3p} where the $K_{\pi 3}$ background is shown as a function of $E/p$ with $\chi^2_{3\pi}>16$ (left) and as a function of $1/\chi^2_{3\pi}$ (right).

In order to diminish this background further, a neural network algorithm was applied \cite{litov}.  A 3-layer neural network was trained on cleanly selected $K_{e3}$ and $K_{\pi 3}$ data samples to distinguish pion and electron electromagnetic showers in the liquid Krypton calorimeter.
Both the $K_{\pi 3}$ and the $K_{e3}$ samples were taken during the same run period as the signal events.
The algorithm used geometric characteristics of showers and tracks and $E/p$ of tracks on the input and returned a control variable which was around 0 for pions and around 1 for electrons.

The background from $K_{\pi 3}$ was estimated as a function of $\chi^2_{3\pi}$ by extrapolating the pion tail
shape under the electron $E/p$ peak as determined from data. 
This was done without the use of the neural network, since the background would be too small otherwise.
Then, with the neural network operating, the previously determined dependence on $\chi^2_{3\pi}$ was fit to the data. 
The background of charged pions misidentified as electrons in a pure $K_{\pi 3}$ sample amounted to 1.2\% and 1.1\% for the KE4 and ETOT triggers, respectively.

After applying all selection criteria, the data samples amounted to 2089 and 3375 events for the KE4 and ETOT triggers, respectively.
The estimated numbers of background events were $25\pm 10$ and $37\pm 5$, respectively.
In total, the sample of $K_{e4}$ events amounted to $5464$ events with 62 background events.

\section{Monte Carlo simulation}

Monte Carlo simulations were used for:
\bi
\item Determination of the acceptance for evaluation of the $K_{e4}$ branching ratio and form factors,
\item Estimation of the background,
\item Estimation of radiative corrections.
\ei
The simulation code is based on event generators for neutral kaon decay channels and full GEANT simulation \cite{geant} of all electromagnetic processes in the NA48 detector, including cascades in the calorimeter.
Samples generated included: more than 1 million $K_{e4}$ events using the form of the decay matrix element as calculated by Pais and Treiman \cite{treiman}, 0.5 million $K_{e3+2\gamma}$ events where one photon was from inner bremsstrahlung and another one from the bremsstrahlung of a charged particle in the detector's material, and 0.5 million $K_{\pi 3}$ events.
Wherever possible, the data were used for background studies in order to be independent of Monte Carlo simulation.

\section{Branching ratio}

The branching ratio of the $K_{e4}$ channel was determined by normalizing it to the $K_{\pi 3}$ channel:
\be
Br(K_{e4})=\frac{N(K_{e4})}{N(K_{\pi 3})}\cdot\frac{a(K_{\pi 3})}{a(K_{e4})}\cdot\frac{\varepsilon(K_{\pi 3})}{\varepsilon(K_{e4})}\cdot Br(K_{\pi 3})
\label{eq3}
\ee
where $N$ stands for the overall number of accepted events, properly corrected for downscaling, $a$ for acceptance, and $\varepsilon$ for trigger efficiency.
For the branching ratio of the reference channel, the value $Br(K_{\pi 3})=(12.58\pm 0.19)\,\%$ was used \cite{pdf}.
The reference sample of $K_{\pi 3}$ was selected using similar cuts 1-11 as for the $K_{e4}$ but requiring $\chi^2_{3\pi}<5$ in cut 9 and no cut on $E/p$.

The acceptances, as calculated from the ratios of accepted to generated Monte Carlo events, are equal to $(3.610\pm 0.017)\,\%$ and $(5.552\pm 0.033)\,\%$ for the $K_{e4}$ and $K_{\pi 3}$ channels, respectively.

From this, the branching ratios for the KE4 and ETOT trigger samples were found to be $(5.30\pm 0.12_{stat}\pm 0.11_{syst})\times 10^{-5}$ and $(5.15\pm 0.09_{stat}\pm 0.12_{syst})\times 10^{-5}$, respectively, and the overall branching ratio is equal to
\be
Br(K_{e4})& = & (5.21\pm 0.07_{stat}\pm 0.09_{syst})\times 10^{-5}
\label{eq4}
\ee
where contributions to the systematic error, in units of 10$^{-5}$, are as follows:

\vspace{3mm}

\begin{tabular}{lc}
$K_{\pi 3}$ branching ratio: & 0.079 \\
$K_{e4}$ form factors: & 0.021 \\
Background from $K_{e3+2\gamma}$: & 0.019 \\
Background from $K_{e4+\gamma}$: & 0.011 \\
Monte Carlo statistics for $K_{e4}$: & 0.024 \\
Monte Carlo statistics for $K_{\pi 3}$: & 0.030 \\
Trigger efficiencies: & 0.005 \\
Background from $K_{\pi 3}$: & 0.001
\end{tabular}

\vspace{3mm}

These systematic errors for branching ratios for the KE4 and ETOT triggers were determined with independent sets of data and Monte Carlo. 
The branching ratio includes radiative events $K_L\rightarrow\pi^0\pi^{\pm}e^{\mp}\nu_e(\bar\nu_e)\gamma$ (called $K_{e4+\gamma}$) left in the sample after all cuts.
Their contribution is accounted for in the systematic error and was estimated using the ratio of accepted $K_{e4}$ to $K_{e4+\gamma}$, known from the Monte Carlo radiative event generator PHOTOS \cite{was}.
The ratio of decay rates can be calculated using formulae of ref. \cite{was}.
It was found that the fraction of all $K_{e4+\gamma}$ events in the final sample, including those with the radiative photon undetected due to acceptance or cuts, is $0.89\pm 0.02\,\%$, in agreement with numbers cited in ref. \cite{pislak}.
For the overall sample this corresponds to 48 events or a $0.042\times 10^{-5}$ contribution to the systematic error.
The systematic uncertainty due to radiative corrections was estimated as $\pm 25\%$ of the maximal contribution to the branching ratio coming from the remaining $K_{e4+\gamma}$ background.
This uncertainty represents an upper bound based on our experience with other decays.

Clearly, the systematic uncertainty is dominated by the error in the $K_{\pi 3}$ branching ratio.

Consistent results were obtained in two, independent analyses.

\begin{figure}                                                  
\begin{center}                                                                                      
\begin{tabular}{cc}
\includegraphics[width=70mm,height=65mm]{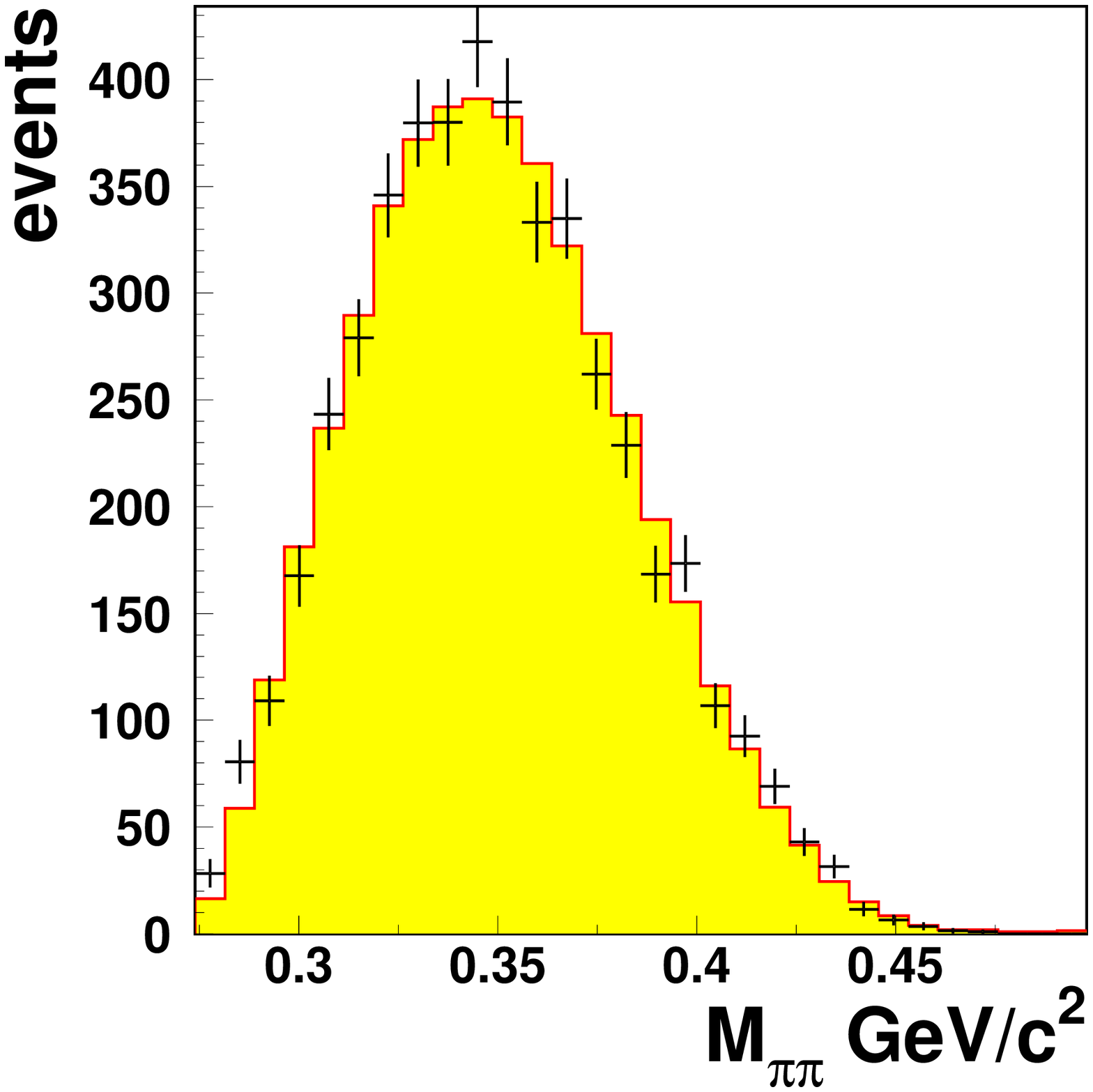} & \includegraphics[width=70mm,height=65mm]{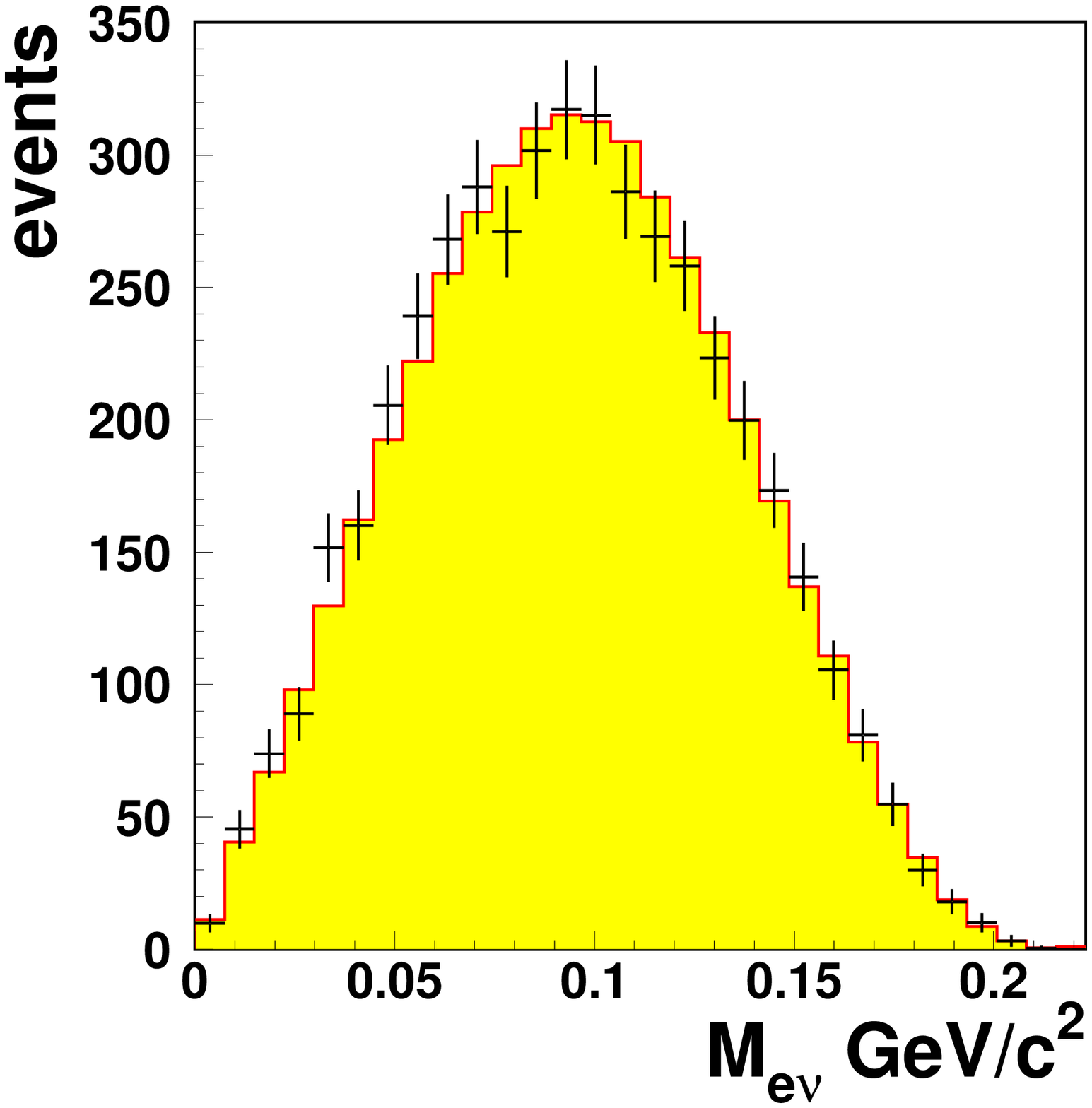} \\
\includegraphics[width=70mm,height=65mm]{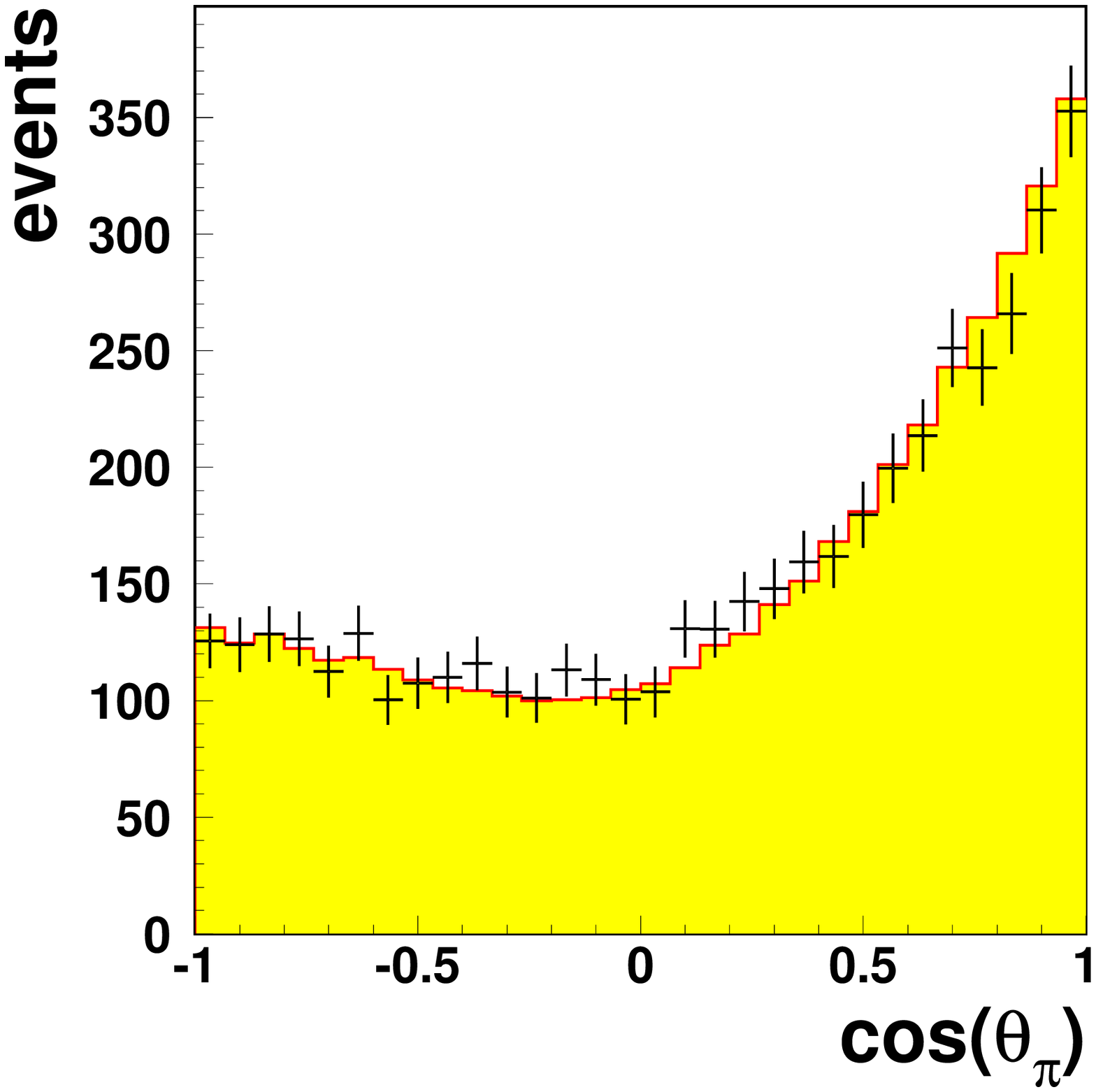} & \includegraphics[width=70mm,height=65mm]{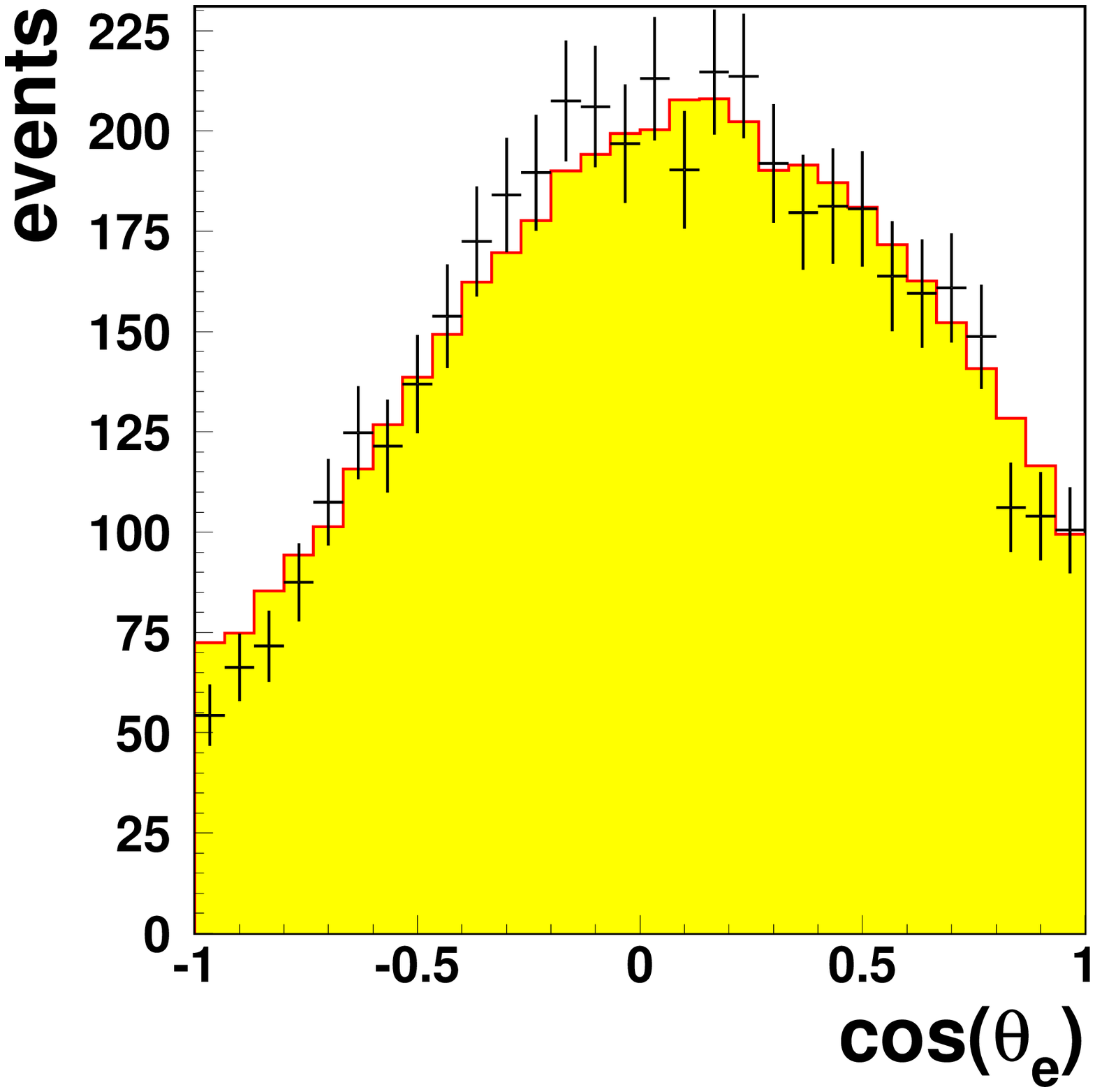} \\
\includegraphics[width=70mm,height=65mm]{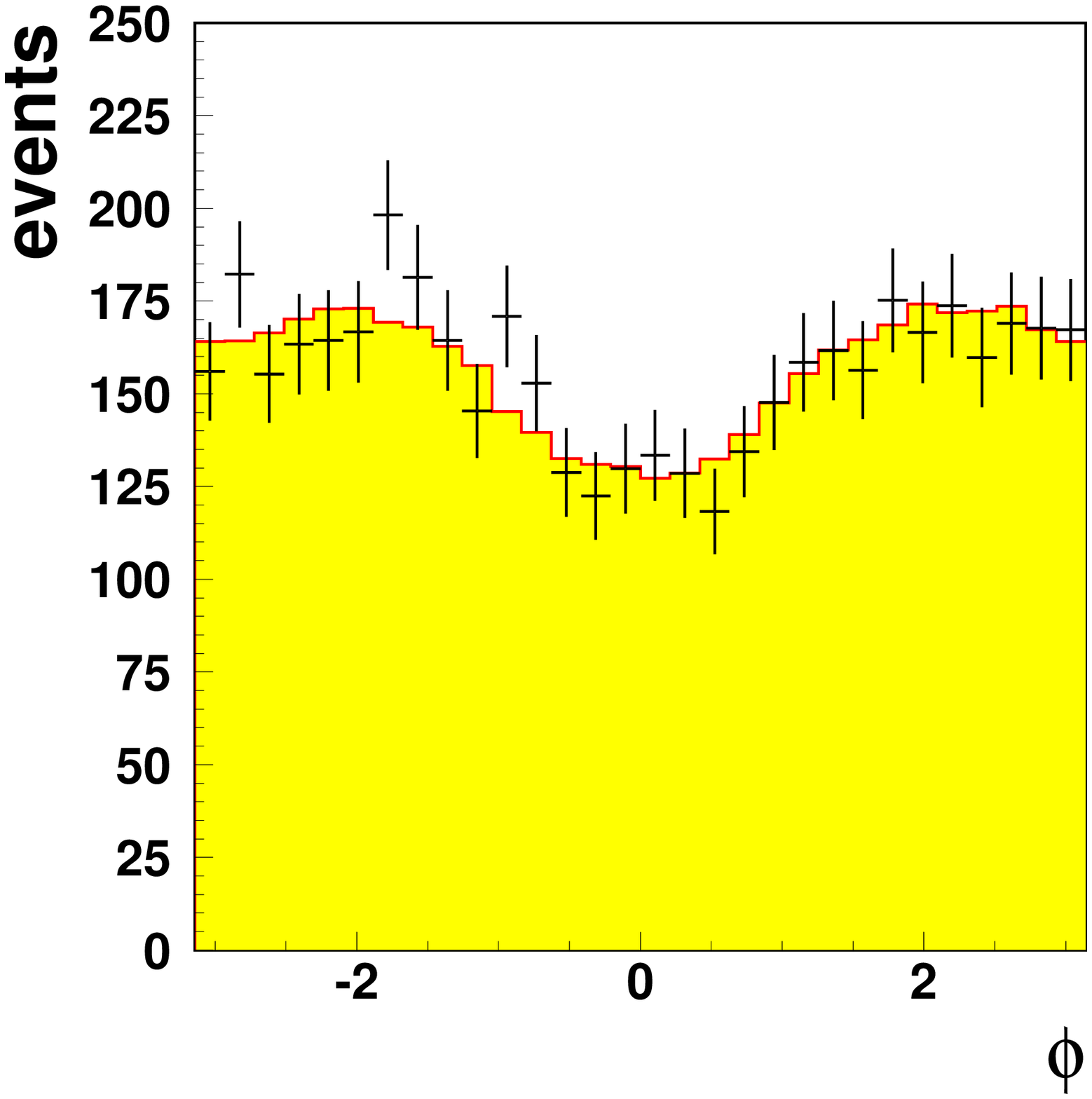} & 
\end{tabular}
\caption{Distributions of the Cabibbo-Maksymowicz variables: $M_{\pi\pi}$ (upper left), $M_{e\nu}$ (upper right), $\cos \theta_{\pi}$ (middle left), $\cos \theta_e$ (middle right) and $\phi$ (bottom) for data from both KE4 and ETOT triggers (points with error bars) with fits (histograms). Acceptance was accounted for in the fit.}
\label{fig4}
\end{center} 
\end{figure}

\section{The form factors}
The form factors were estimated by fitting the differential distributions, as found in the analytic expression in ref. \cite{rosselet}, to the empirical distributions of the C-M variables.
The detector acceptance was accounted for by multiplying the theoretical function by the distribution of accepted events generated by Monte Carlo with flat form factors.
In order to account for $K_{e4+\gamma}$ events, the radiative generator PHOTOS \cite{was} was used.
We generated radiative events with one bremsstrahlung photon emitted by a charged particle in the final state.

Since the data sample of $K_{e4}$ events was not large enough to allow a five-dimensional analysis, a simultaneous fit to all one-dimensional projections was performed.
It was found, and checked with Monte Carlo, that the maximum correlation coefficient between points on projections depends on the number of bins $n$ as ${\cal O}(1/n^2)$ and therefore can be neglected for $n\gtrsim 10$.
The theoretical curves were found for each projection by integrating over the remaining four variables.
The method was checked by fitting the two-dimensional distributions of $M_{\pi\pi}$ vs. $M_{e\nu}$ and $\theta_{\pi}$ vs. $\theta_e$, after integration over the remaining three variables, and finding results in satisfactory agreement with a simultaneous one-dimensional fit.
The one-dimensional distributions of the C-M variables, experimental and fits, are presented in figs \ref{fig4}.
The values of the form factors were found from the fit to data from combined KE4 and ETOT triggers to be:
\vspace{3mm}
\be
\bar{f}_s & = & 0.052 \pm 0.006_{stat}\pm 0.002_{syst} \nn \\
\bar{f}_p & = & -0.051 \pm 0.011_{stat} \pm 0.005_{syst} \nn \\
\lambda_g & = & 0.087 \pm 0.019_{stat} \pm 0.006_{syst} \nn \\
\bar{h} & = & -0.32 \pm 0.12_{stat} \pm 0.07_{syst}
\ee
with $\chi^2/ndf=137/146$.

The phases $\delta_s$ and $\delta_p$ were not determined in this analysis.
The factor $g(0)$ can be determined in a model-dependent way, using the branching ratio and predictions of CHPT \cite{widhalm}.

The systematic errors are dominated by the background, with minor contributions from Monte Carlo statistics and the shapes of the $E/p$ and $p_T$ distributions used for background subtractions.
As with the branching ratio, the contribution from the radiative $K_{e4+\gamma}$ events to the systematic errors was estimated as $\pm 25\%$ of the difference between the form factors calculated with and without radiative events in the Monte Carlo background.
As a cross check, we fitted separately the KE4 and ETOT trigger samples obtaining results consistent within the statistical errors.
Additional checks were performed by inspecting ratios of C-M variable distributions between trigger samples, and between data and MC weighted by fitted form factors, and no significant discrepancies were found. 
The stability of the results was also examined by increasing and decreasing the amount of the $K_{\pi 3}$ background through varying the $\chi^2_{3\pi}$ cut and found to be satisfactory.
Also the fitting procedure was checked against background addition.
Radiative corrections affect the values of form factors within one standard deviation of the statistical error.

Assuming the $K_{e4}$ hypothesis, there are two kinematically allowed solutions for the kaon energy and hence the neutrino energy.
The choice has no effect on $M_{\pi\pi}$ but can slightly affect other C-M variables.
Both solutions with equal weights of 0.5 were used for the form factor fits.
The systematic effect of making the wrong choice was examined using Monte Carlo and found to be negligible.

All steps of the analysis, viz. event selections, backgrounds and fits, were done twice and independently, and the results were in good agreement.

\section{Discussion and conclusions}

\begin{figure}
\begin{center}
\begin{tabular}{cc}
\includegraphics[width=100mm,height=100mm]{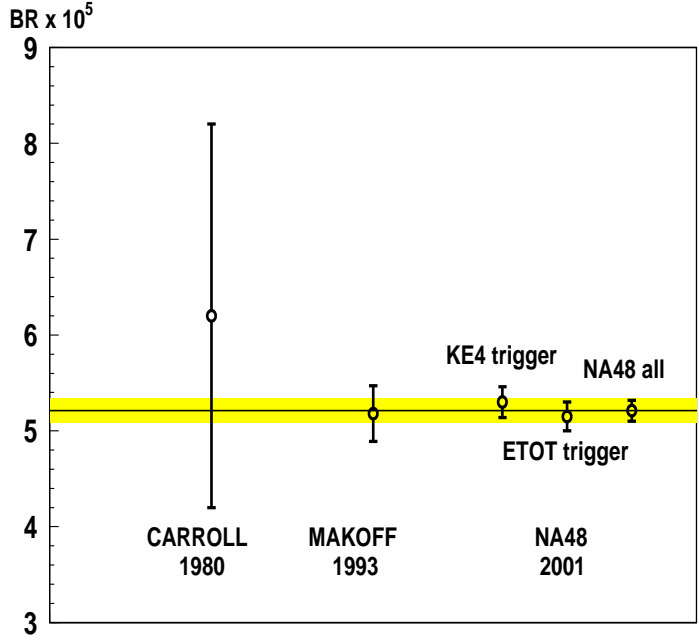}
\end{tabular}
\caption{$K_{e4}$ branching ratios from refs \cite{carrol} and \cite{makoff} and from the present work (results from data using both triggers are shown). The shaded belt corresponds to the overall NA48 result.}
\label{fig5}
\end{center}                                                                                                 \end{figure}

The $K_{e4}$ branching ratio measured by NA48 is consistent with previous measurements \cite{carrol,makoff} within errors and is more accurate by a factor of 2.5 than that of ref. \cite{makoff}, both statistically and systematically (cf. fig. \ref{fig5}).
The form factors $\bar{f}_p$, $\lambda_g$ and $\bar{h}$ have also significantly higher accuracy and agree within errors with ref. \cite{makoff}, whereas the value of $\bar{f}_s$ differs by two standard deviations.

We found a non-zero value of $\bar{f}_s$, allowing for the violation of the $\Delta I=1/2$ rule at the percent level.
As discussed in ref. \cite{berends}, admixtures of $\Delta I=3/2$ and $\Delta I=5/2$ to the $K_{e4}$ decay amplitude, at the level comparable to that of the $K\rightarrow\pi\pi$ decays, can be expected.

We note good agreement of our $\bar h$ value with previous neutral and charged $K_{e4}$ studies \cite{rosselet,pislak,makoff} and with the theoretical prediction \cite{bijnens}, essentially independent of the coefficients $L_i$ of the chiral Lagrangian.

The $K_{e4}$ decay is helpful in determining of the chiral coupling parameter
$L_3$, which attracts theoretical interest, extending beyond CHPT, for its direct relation to the gluon condensate and the constituent quark mass \cite{bijnens,espriu}.
The neutral $K_{e4}$ branching ratio is mainly sensitive to $L_3$ and very little to $L_5$ and $L_9$ \cite{widhalm}.
Using this dependence one gets
\be
L_3=(-4.1\pm 0.2)\times 10^{-3} 
\label{eq6}
\ee
Also, in CHPT the form factor $\bar{f}_p$ depends linearly on $L_3$ with directly computable numerical constants.
In addition, the form factor $\lambda_g$ depends linearly on $L_3$ with numerical constants dependent on the well known pion decay constant $F_{\pi}$.
Using either $\bar{f}_p$ or $\lambda_g$, we get values for $L_3$ consistent within errors with the value of eqn. (\ref{eq6}) but with five times larger uncertainties.
The value of $L_3$ determined in this work is more accurate than theoretical estimates from CHPT fits based on previously available data \cite{bijnens}.
Our result is also compatible with the result of ref. \cite{makoff}.

\section{Acknowledgements}

It is a pleasure to thank the technical staff of the participating laboratories, universities and affiliated computing centres for their efforts in the construction of the NA48 apparatus, in the operation of the experiment, and in the processing of the data.
We also thank Gerhard Ecker for discussing theoretical issues related to the $K_{e4}$ decay.

\end{document}